\documentclass[apjl]{emulateapj}
\usepackage[dvips]{color}

\def\ltsima{$\;\buildrel < \over \sim \;$}
\def\simlt{\lower.5ex \hbox{\ltsima}}
\def\gtsima{$\;\buildrel > \over \sim \;$}
\def\simgt{\lower.5ex \hbox{\gtsima}}

\shorttitle{Water vapor around carbon-rich AGB stars}
\shortauthors{Neufeld et al.}

\begin{document}

\title{The widespread occurence of water vapor in the circumstellar
envelopes of carbon-rich AGB stars:
first results from a survey with Herschel$^*$/HIFI}
\author{D.~A. Neufeld\altaffilmark{1},
E.~Gonz\'alez-Alfonso\altaffilmark{2},
G.~Melnick\altaffilmark{3}, 
R.~Szczerba\altaffilmark{4},
M.~Schmidt\altaffilmark{4},
L.~Decin\altaffilmark{5,6},
J.~Alcolea\altaffilmark{7},
A.~de~Koter\altaffilmark{6,8}
F.~L.~Sch\"oier\altaffilmark{9},
V.~Bujarrabal\altaffilmark{10},
J.~Cernicharo\altaffilmark{11},
C.~Dominik \altaffilmark{6,12},
K.~Justtanont\altaffilmark{9},
A.~P.~Marston\altaffilmark{13},
K.~Menten\altaffilmark{14},
H.~Olofsson\altaffilmark{9,15},
P.~Planesas\altaffilmark{10,16},
D.~Teyssier\altaffilmark{11} and
L.~B.~F.~M.~Waters\altaffilmark{17,6}}

\altaffiltext{*}{Herschel is an ESA space observatory with science instruments provided
by European-led Principal Investigator consortia and with important
participation from NASA}
\altaffiltext{1}{The Johns Hopkins University, Baltimore, MD 21218, USA} 
\altaffiltext{2}{Departamento de F\'{\i}sica, Universidad de Alcal\'a de Henares, Campus Universitario, E-28871 Alcal\'a de Henares, Madrid, Spain} 
\altaffiltext{3}{Harvard-Smithsonian CfA, Cambridge, MA 02138, USA} 
\altaffiltext{4}{N. Copernicus Astronomical Center, Toru{\'n}, Poland} 
\altaffiltext{5}{Instituut voor Sterrenkunde, K.~U. Leuven, Belgium} 
\altaffiltext{6}{Sterrenkundig Instituut Anton Pannekoek, University of Amsterdam, The Netherlands} 
\altaffiltext{7}{Observatorio Astron\'omico Nacional (IGN), Madrid, Spain}
\altaffiltext{8}{Astronomical Institute, Utrecht University, The Netherlands}
\altaffiltext{9}{Onsala Space Observatory, Dept. of Radio and Space Science, Chalmers  University of Technology, SE--43992 Onsala, Sweden} 
\altaffiltext{10}{Observatorio Astron\'omico Nacional (IGN), Alcal\'a de Henares, Spain} 
\altaffiltext{11}{CAB, INTA-CSIC, Ctra de Torrej\'on a Ajalvir, Madrid, Spain} 
\altaffiltext{12}{Department of Astrophysics/IMAPP, Radboud University Nijmegen, The Netherlands} 
\altaffiltext{13}{European Space Astronomy Centre, Villanueva de la Ca\~nada, Madrid, Spain} 
\altaffiltext{14}{MPI f{\"u}r Radioastronomie, Bonn, Germany} 
\altaffiltext{15}{Department of Astronomy,  Stockholm University, Sweden} 
\altaffiltext{16}{Joint ALMA Observatory, Santiago, Chile} 
\altaffiltext{17}{SRON Netherlands Institute for Space Research, Utrecht, The Netherlands}

\begin{abstract}

We report the preliminary results of a survey for water vapor in a sample of eight C stars with large mid-IR continuum fluxes: V384 Per, CIT 6, V Hya, Y CVn, IRAS 15194-5115, V~Cyg, S~Cep, and IRC+40540.  
This survey, performed using the { HIFI} instrument on board the {\it Herschel Space Observatory}, entailed observations of the lowest transitions of both ortho- and para-water: the 556.936~GHz $1_{10}-1_{01}$ and 1113.343~GHz $1_{11}-0_{00}$ transitions, respectively.  Water vapor was unequivocally detected in all eight of the target stars.  Prior to this survey, IRC+10216 was the only carbon-rich AGB star from which thermal water emissions had been discovered, in that case with the use of the {\it Submillimeter Wave Astronomy Satellite (SWAS)}.  Our results indicate that IRC+10216 is not unusual, except insofar as its proximity to Earth leads to a large line flux that was detectable with {\it SWAS}.  The water spectral line widths are typically similar to those of CO rotational lines, arguing against the vaporization of a Kuiper belt analog (Ford \& Neufeld 2001) being the {\it general} explanation for water vapor in carbon-rich AGB stars.  
There is no apparent correlation between the ratio of the integrated water line fluxes to the 6.3 micron continuum flux -- a ratio which measures the water outflow rate -- and the total mass-loss rate for the stars in our sample.

\end{abstract}

\keywords{circumstellar matter --- stars: AGB and post-AGB --- stars: abundances --- submillimeter: stars  }

\section{Introduction}

Over the past two decades, observations of the envelopes of AGB (asymptotic giant branch) stars have revealed a molecular composition that is still not entirely understood.
In early models for the chemistry within the circumstellar envelopes of AGB stars, the inner envelope was assumed to possess a composition determined by thermochemical equilibrium (TE) within the stellar photosphere, while the outer envelope, exposed to the interstellar radiation field, exhibits a time-dependent gas-phase chemistry that is driven by the photodissociative effects of ultraviolet radiation (e.g.\ Glassgold 1996).  In TE, the carbon-to-oxygen ratio is crucial in determining the photospheric composition; cool oxygen-rich stars, with $\rm C/O < 1$, are expected to have photospheres that are dominated by CO and H$_2$O, while those of carbon-rich stars will be dominated by CO and C$_2$H$_2$.   However, a variety of observations have revealed anomalous abundances for several molecules that cannot be explained by models in which the inner envelope entirely reflects the photospheric abundances in TE and the ensuing chemistry proceeds in the gas phase.  More recent models involving shocks (Willacy \& Cherchneff 1998; Cherchneff 2006), grain surface reactions (e.g. Willacy 2004), the vaporization of orbiting objects (Ford \& Neufeld 2001), or the penetration of UV radiation through a clumpy outflow to the inner envelope (Decin et al.\ 2010) have been variously invoked to explain the observed abundances of such species as CO, HCN, SiO, CS, H$_2$CO, OH and H$_2$O in the envelopes of both O-rich and C-rich AGB stars.

\begin{deluxetable*}{lcrccrcccccc}

\tablewidth{0pt}
\tabletypesize{\scriptsize}
\tablecaption{Sample of C stars surveyed for water vapor} 
\tablehead{Star & R.A. & Dec.    & D  & $\dot M$  & $v_{\rm sys}$ & \multicolumn{2}{c}{Date (2010)} & \multicolumn{2}{c}{Duration (s)} & \multicolumn{2}{c}{Noise (mK) $^f$}\\
& (J2000) & (J2000) &  (pc) & (a) & (b) & 557 & 1113   & 557 & 1113 & 557 & 1113  \\
&&&&&&  GHz & GHz & GHz & GHz  & GHz & GHz  } 

\startdata
V384 Per & 03 26 29.5 &  +47 31 50 & 740 & $5.5$ & --16.2 &  Sep 1  & Sep 2 &  3972$^d$ &  987 & 2.4 & 14\\ 
IRC+10216$^c$ & 09 47 57.4 & +13 16 44 & 110 & $15$ & --25.5 & May 4,11 & May 11 & 3150$^d$ & 3200$^a$ & 2.7 & 14 \\ 
CIT 6$^c$ & 10 16 02.2 & +30 34 19 & 320 & $1.9$ & --1.6 &  May 11 & May 11 & 4719\phantom{$^d$} & 987 & 1.9 & 16 \\ 
V Hya & 10 51 37.2 & --21 15 00 & 600 & $ 2.1$ & --17.0 &  Jun 23 & Jun 9  & 3972$^d$ & 987 & 3.4 & 15 \\ 
Y CVn & 12 45 07.8 & +45 26 25 & 220 & $0.15$ & 21.1 & Jun 23 & Jun 9 & 3972$^d$ & 987 & 2.2 & 16 \\ 
IRAS 15194-5115 & 15 23 04.9 & --51 25 59 & 590 & $18$ & --15.0 & Aug 2 & N/A$^e$  & 3972$^d$ & N/A$^e$ & 2.1 & N/A$^e$\\
V Cyg & 20 41 18.3 & +48 08 29 & 530 & $2.5$ & 14.3 & Jun 23 & Mar 5 & 3972$^d$ & 987 & 2.6 & 16 \\ 
S Cep & 21 35 12.8 & +78 37 28 & 720 & $3.4$ & --15.5 & N/A$^e$ & May 12 & N/A$^e$ & 987 & N/A$^e$ & 13 \\ 
IRC+40540$^c$  & 23 34 27.7 & +43 33 02 & 780 & $7.0$ & --17.0 & Jun 23 & Jun 9 & 3972$^d$ & 987 & 2.5 & 15\\ 
\enddata

\tablenotetext{a}{Mass loss rate in units of $10^{-6}\,M_\odot\,\rm yr^{-1}$}
\tablenotetext{b}{Systemic velocity  in km s$^{-1}$ relative to the local standard of rest}
\tablenotetext{c}{IRC+10216, CIT6, and IRC+40540 are known also as CW Leo, RW LMi, and LP And respectively.}
\tablenotetext{d}{Divided into two observations of equal duration with slightly different LO settings (see text)}
\tablenotetext{e}{N/A = not available; planned observation not yet carried out}
\tablenotetext{f}{R.m.s.\ in a 1 km s$^{-1}$ bandwidth}

\end{deluxetable*}  

While most of the principal circumstellar molecules are easily detected by means of ground-based observations, water has been more elusive.  Although water maser emissions from oxygen-rich stars can be detected from the ground, thermal water emissions are detectable only with satellite observatories.  As expected, the {\it Infrared Space Observatory (ISO)} detected luminous water emissions from a variety of oxygen-rich AGB stars (e.g. Barlow 1999, Neufeld et al.\ 1999).  More surprisingly, later observations performed (Melnick et al.\ 2001) with the {\it Submillimeter Wave Astronomy Satellite (SWAS)} -- and subsequently confirmed with the ODIN satellite (Hasegawa et al.\ 2006) -- have led to the discovery of water vapor in the envelope of the nearest carbon-rich AGB star { with a massive envelope}, IRC+10216.  Prior to the recent launch of the {\it Herschel Space Observatory} (Pilbratt et al.\ 2010),  this detection of the lowest, $1_{10} - 1_{01}$, rotational transition of ortho-water at 556.936 GHz by SWAS and ODIN remained the only case in which thermal water vapor emission had been detected from a C star.

The {\it Herschel Space Observatory} presents an opportunity to search for water vapor in C stars with unprecedented sensitivity; with a primary mirror of collecting area $\sim 30$ times that of the SWAS mirror, and with its Heterodyne Instrument for the Far Infrared (HIFI; de Graauw et al. \ 2010) employing cryogenic mixers of much lower noise temperature than those flown on SWAS, {\it Herschel} makes it { feasible} to search for water in AGB stars considerably more distant than IRC+10216.  In addition, {\it Herschel} provides access to multiple transitions of water vapor.  As part of the HIFISTARS Key Program, we carried out a survey for water vapor in a sample comprising eight C stars in addition to IRC+10216 --  V384 Per, CIT 6, V Hya, Y CVn, IRAS 15194-5115, V Cyg, S Cep, and IRC+40540 -- with the goal of determining whether IRC+10216 is unusual (in addition to being nearby), or whether water vapor is widely detectable in the envelopes of C stars.  
Our survey involves observations of the lowest transitions of both ortho- and para-water, the 556.936~GHz $1_{10}-1_{01}$ and 1113.343~GHz $1_{11}-0_{00}$ transitions respectively, toward a selection of C stars with large mid-IR continuum fluxes.  Table 1 lists the sources in our sample, together with estimates of their distances, $D$, mass-loss rates, ${\dot M}$, and systemic velocities relative to the local standard of rest.  
{ Although the quantity ${\dot M}/D^2$ is fairly well constrained, uncertainties in $D$ make both parameters quite uncertain} -- a fact that is demonstrated by the wide dispersion of estimates given in the literature -- but in the interest of uniformity we adopt the values tabulated by Groenewegen et al.\ (2002), supplemented by those of Sch\"oier \& Olafsson (2001) for the case of Y CVn (for which estimates were not given by 
Groenewegen et al.\ 2002).  With the exception of Y CVn, which is classified as a J-type variable carbon star of spectral type CV5J (with V indicating variability and J indicating a large $\rm ^{13}C/^{12}C$ ratio), all the stars in the sample are of spectral type CV6 or CV7.  V Hya is also of a somewhat different character than the other sources; { it is a semi-regular variable (SRa), quite probably a binary (Sahai et al.\ 2003), and possesses} a high velocity outflow and bipolar symmetry that suggests it is in an early stage of {\it post}-AGB evolution (Kahane et al.\ 1996, Knapp et al.\ 1997).  { The mass-loss rate given in Table 1 is for the component with outflow speed $\sim 15\,\rm km\,s^{-1}$ (identified as the ``normal slow wind'' by Knapp et al.\ 1997).}

In this {\it Letter}, we report preliminary results of this survey, based on the observations performed to date toward the eight target stars.  The detection of the $1_{11}-0_{00}$ water transition toward V Cygni in the very first observation performed in this survey has been reported previously (Neufeld et al.\ 2010).  The data acquisition and reduction are described in \S 2, and the results are presented in \S 3.  In \S 4, the results are discussed with reference to various scenarios that have been proposed for the origin for water vapor around C stars.

\begin{figure}
\includegraphics[scale=0.9]{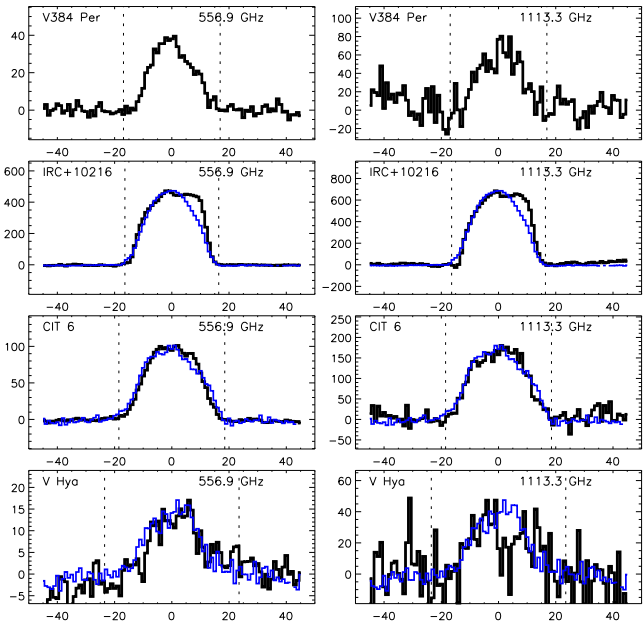}

\caption{Black histogram: Herschel/HIFI spectra of the $1_{10}-1_{01}$ 556.936~GHz transition of ortho-H$_2$O (left) and the  $1_{11}-0_{00}$ 1113.343~GHz transition of para-H$_2$O (right) observed toward V384 Per, IRC+10216, CIT6, and V Hya.  Blue histogram: analogous spectra for CO $J= 10-9$, where available, scaled to the same peak antenna temperature.  The spectra have been continuum-subtracted and rebinned to a spectral resolution of 1~km~s$^{-1}$.
Doppler velocities are expressed in km~s$^{-1}$ relative to the systemic velocity of the source, and antenna temperatures (vertical axis) are shown in mK. Dotted lines indicate expansion velocities from the literature ({ Groenewegen et al.\ 2002).}}
\end{figure}

\begin{figure}
\includegraphics[scale=0.9]{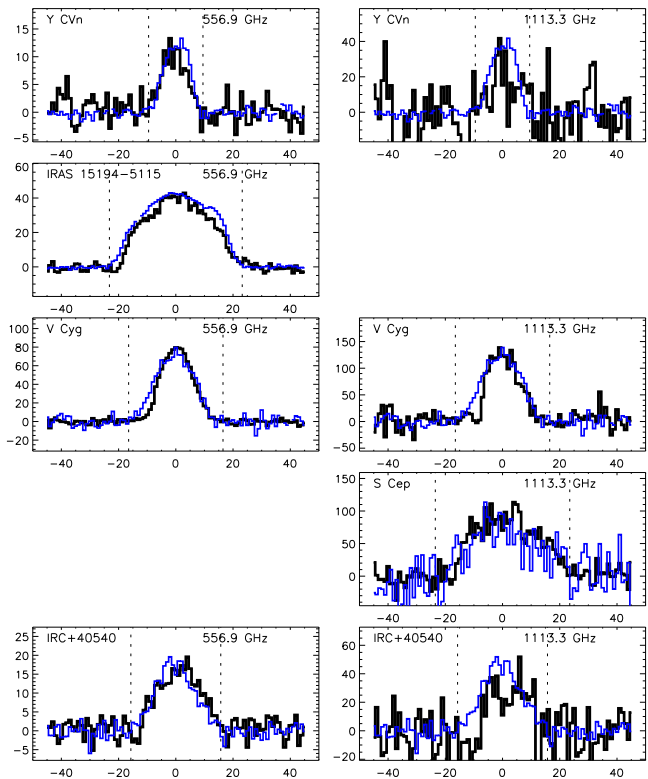}
\caption{Same as Fig. 1, for Y CVn, IRAS-15194, V Cyg, S Cep and IRC+40540.  For Y CVn and IRAS-15194 the blue histogram shows CO $J=6-5$ instead of $J=10-9$}
\end{figure}

\section{Observations and data reduction}

The observations reported here were carried out during the period March -- September 2010 as part of the HIFISTARS Key Program.  We used the HIFI instrument in dual beam switch (DBS) mode to observe two rotational transitions of water vapor towards each source listed in Table 1.  The 556.936~GHz $1_{10}-1_{01}$ and 1113.343~GHz $1_{11}-0_{00}$ transitions were targeted respectively in lower and upper sidebands of Bands 1b and 4b of the HIFI instrument.  The telescope beam was centered at the coordinates listed in Table 1, with the reference positions located at offsets of 3$^\prime$ on either side of each source.  The beam sizes were 38$^{\prime \prime}$ and 19$^{\prime \prime}$ (half power beam width) respectively for the 556.936~GHz and 1113.343~GHz transitions.  The data were acquired using the Wide Band Spectrometer (WBS), which provides a spectral resolution of 1.1~MHz and a simultaneous bandwidth of $\sim 4$~GHz.  For each observation, the date and total duration are given in Table 1, { together with the r.m.s.\ noise achieved in a 1 km s$^{-1}$ bandwidth.}
As discussed in Neufeld et al.\ (2011), the data were processed using the standard HIFI pipeline to Level 2, providing fully calibrated spectra with the intensities expressed as antenna temperature and the frequencies in the frame of the Local Standard of Rest (LSR).
For several observations, we used a pair of local oscillator (LO) frequencies, separated by a small offset, to confirm the assignment of any observed spectral feature to either the upper or lower sideband of the (double side band) HIFI receivers.  The resultant spectra were coadded so as to recover the signal-to-noise ratio that would have been obtained at a single LO setting.  Spectra obtained for the horizontal and vertical polarizations were found to be very similar in their appearance and noise characteristics and were likewise coadded. 





\section{Results}

Figures 1 and 2 show the continuum-subtracted spectra obtained from  each of the observations listed in Table 1, with the water spectra plotted in black.  To improve the signal-to-noise ratio, each spectrum has been rebinned to a spectral resolution of 1~km~s$^{-1}$.  In each case, the velocity scale is relative to the systemic velocity of the source (Table 1, as given by Knapp et al.\ 1998 for Y CVn and Groenwegen et al.\ 2002 for the other stars.)  The blue histograms show, for comparison, the CO $J= 6-5$ ({ in the case of} IRAS 15194-5115 and Y CVn) or CO $J= 10-9$ (other stars) spectra obtained with HIFI, scaled to have the same peak antenna temperature as the water spectra.  { No HIFI spectra of CO are yet available for V384 Per.}
Water vapor is clearly detected in each of the eight target stars by means of one or both of the observed transitions.  As predicted by excitation models that we performed in anticipation of {\it Herschel} (Gonz{\'a}lez-Alfonso, Neufeld, \& Melnick 2007; hereafter GNM), { and given the sensitivity of {\it Herschel} as a function of frequency},
the $1_{10}-1_{01}$ transition provides the most sensitive probe of water in C stars (although in the single case of S Cep, the detection of water vapor was obtained through observations of the $1_{11}-0_{00}$ transition alone, the $1_{10}-1_{01}$ having not yet been observed.)  In Table 2, we list the integrated antenna temperatures and line fluxes for each transition, together with the $6.3\mu$m continuum flux obtained from the {\it ISO} observations of Sloan et al.\ (2003) or -- { in the cases of CIT6, V Hya and IRAS 15194-5115} -- estimated from the 8.8$\mu$m fluxes reported by Monnier, Geballe \& Danchi (1998).  { The absolute flux calibration for ISO spectroscopy was accurate to within $\sim 30 \%$.}
The continuum flux is of particular interest because theoretical models for the excitation of water in C stars (Ag{\'u}ndez \& Cernicharo 2006; GNM) indicate that the observed $1_{10}-1_{01}$ and $1_{11}-0_{00}$ rotational emissions are excited primarily by radiative pumping through the $\nu_2$ vibrational band at 6.3 $\rm \mu m$.

\begin{deluxetable}{lrrrrr}

\tablewidth{0pt}
\tablecaption{Water line intensities and fluxes}
\tablehead{Star & \multicolumn{2}{c}{$\int{T_A dv}$}  & \multicolumn{2}{c}{Line flux$^a$} & $F_{6.3}$\\
                 & \multicolumn{2}{c}{( K km~s$^{-1}$ )} & \multicolumn{2}{c}{(10$^{-17}$ W m$^{-2})$} & ($10^3 \, \rm Jy$) \\ 
& 557 & 1113   & 557 & 1113  & \\
&  GHz & GHz & GHz & GHz   &}

\startdata
V384 Per 	& 0.64 & 1.12 & 0.55 & 1.95 & 0.46\\
IRC+10216  	&10.3\phantom{0} &14.6\phantom{0} & 8.70 & 25.22 & 22.3 \phantom{0} \\
CIT 6 		& 2.17 & 3.81 & 1.86 & 6.62 & 4.00 \\
V Hya 		& 0.28 & $0.65$ & 0.24 & 1.13  & 1.70 \\
Y CVn 		& 0.11 & $0.27$ & 0.09 & 0.47 & 0.37 \\
IRAS 15194-5115 & 1.12 & N/A$^b$  & 0.96 & N/A$^b$ & 1.00\\
V Cyg 		& 1.10 & 1.74 & 0.94 & 3.02 & 1.04 \\
S Cep 		& N/A$^b$  & 2.54 & N/A$^b$ & 4.41 & 0.94 \\
IRC+40540	& 0.29 & $0.37$ & 0.25 & 0.64 & 0.40
\enddata

\tablenotetext{a}{We conservatively estimate the typical flux uncertainty as better than $15\%$, except for V Hya, Y CVn, and IRC+40540.  In these sources, the 1113 GHz flux uncertainties are dominated by noise; we estimate them as $20\%$, $30\%$, and $25\%$ (1 $\sigma$) respectively.}
\tablenotetext{b}{Planned observation not yet carried out}

\end{deluxetable}

\section{Discussion}

Although there are real differences between the H$_2$O and CO line profiles for some of the spectra shown in Figures 1 and 2 (most notably for the IRC+10216 spectra which have the highest signal-to-noise ratio), the line widths are typically quite similar.  In particular, the half width at zero intensity is typically close to literature estimates of the expansion velocity (outer dotted vertical lines in each panel).  This feature of the spectra indicates that water-emitting gas is typically present over a wide range of outflow directions, including streamlines directed close to the line-of-sight, and argues strongly against the vaporization of a Kuiper belt analog (Ford \& Neufeld 2001) being the {\it general} explanation for water vapor in carbon-rich AGB stars.  Any collection of orbiting icy objects with a flattened structure (like the Kuiper belt) would release water that would be entrained in the outflow and carried along streamlines with small inclinations to the equatorial plane of the structure.  Viewed nearly edge-on, such an anisotropic distribution of water vapor could indeed yield a spectral line of width comparable to that of the CO lines; but viewed nearly face-on, the water lines would be considerably narrower. Thus, while the vaporization of a Kuiper belt might account for the water line profile observed in any {\it individual} source, the statistical ensemble now available from our survey is not consistent with Kuiper belt vaporization as a general explanation for water vapor.  (Note, however, that based upon multitransition observations available for IRC+10216 but not for the other sources in our sample, Decin et al.\ 2010, and subsequently Neufeld et al.\ 2011, have argued that the distribution of water vapor in that source is inconsistent with the vaporization of small icy objects on circular orbits being the source of the water in that source.)

\begin{figure}
\includegraphics[scale=0.50]{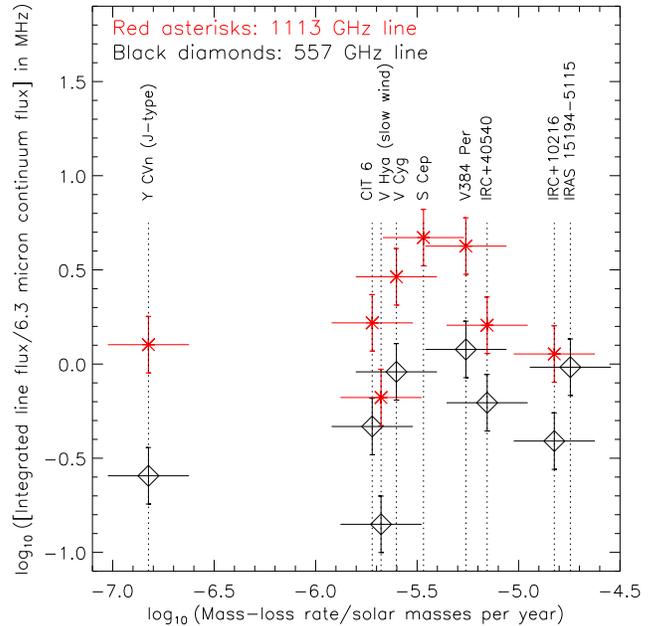}

\caption{Ratio of integrated line flux to 6.3$\mu$m continuum flux for the 557 GHz (black diamonds) and 1113 GHz (red asterisks) water lines, as a function of the estimated total mass-loss rate (Table 1)}
\end{figure}

Detailed calculations by GNM predict luminosities for the $1_{10}-1_{01}$ and $1_{11}-0_{00}$ transitions that are roughly proportional to the 6.3 $\mu$m continuum luminosity, and that the ratio of integrated line flux to 6.3$\mu$m continuum flux is an increasing function of the water mass-loss rate.  Because collisional excitation is negligible except in the densest CSEs, the line fluxes show almost no dependence upon the gas density; thus the ratio of integrated line flux, $F_l$, to continuum flux, $F_{6.3}$, is nearly independent of the total mass-loss rate.  
In Figure 3, we plot the $F_l/F_{6.3}$ ratio for each star as a function of the total mass-loss rate, ${\dot M}$.  Here, we used the results tabulated in Table 2, with integrated line fluxes in units of $10^{-17}$ W m$^{-2}$, and the 6.3 $\mu$m continuum fluxes are in units of $10^3$ Jy; thus the plotted ratio has units of MHz.  
The plotted values of ${\dot M}$ are the literature estimates tabulated in Table 1, { with error bars indicating an estimated uncertainty of 0.2 dex}.  If the water vapor present in these CSEs is produced by chemical means within the outflow with some universal H$_2$O/H$_2$ ratio,  the $F_l/F_{6.3}$ ratio (measuring water outflow rate) and the total mass-loss rate would show a positive correlation.  There is no discernable correlation in Figure 3, suggesting that stars with larger mass outflow rates tend to have smaller water {\it abundances} (although it remains unclear whether this tendency indicates a {\it causative} relationship).  Quantitative estimates of the water outflow rates and abundances will require detailed modeling that will be the subject of a future paper (and that we hope to constrain further with future {\it Herschel} observations.)  For the two sources that have been studied in detail to date, V Cyg (Neufeld et al.\ 2010) and IRC+10216 (GNM; Neufeld et al.\ 2011), the derived water outflow rates support the lack of positive correlation suggested by Figure 3.  While IRC+10216 has a total mass-loss rate $\sim 10$ times that of V Cyg, it shows a water outflow rate that is only $\sim$ one-third as large; { the inferred H$_2$O/H$_2$ abundance ratio in the CSE of V Cyg ($2 \times 10^{-6}$) is therefore $\sim 25$ times larger than that inferred for IRC+10216 ($8 \times 10^{-8}$).}

\acknowledgments

HIFI has been designed and built by a consortium of institutes and university departments from across
Europe, Canada and the United States under the leadership of SRON Netherlands Institute for Space
Research, Groningen, The Netherlands and with major contributions from Germany, France and the US.
Consortium members are: Canada: CSA, U.~Waterloo; France: CESR, LAB, LERMA, IRAM; Germany:
KOSMA, MPIfR, MPS; Ireland, NUI Maynooth; Italy: ASI, IFSI-INAF, Osservatorio Astrofisico di Arcetri-
INAF; Netherlands: SRON, TUD; Poland: CAMK, CBK; Spain: Observatorio Astron\'omico Nacional (IGN),
Centro de Astrobiolog\'a (CSIC-INTA). Sweden: Chalmers University of Technology - MC2, RSS \& GARD;
Onsala Space Observatory; Swedish National Space Board, Stockholm University - Stockholm Observatory;
Switzerland: ETH Zurich, FHNW; USA: Caltech, JPL, NHSC.

This research was performed, in part, through a JPL contract funded by the National Aeronautics and Space Administration.
E.G-A  is a Research Associate at the Harvard-Smithsonian 
Center for Astrophysics.
R.~Sz.\ and M.~Sch.\ acknowledge support from grant N~203~581040.   J.A. and V.B. have been supported by grant ``ASTROMOL" (CSD2009-00038).

\end{document}